# CHAPTER 3

# Structural and Magnetic Properties of $Er_3Fe_{5-x}Al_xO_{12}$ Garnets

Ibrahim Bsoul[1,a], Khaled Hawamdeh[1,b], Sami H. Mahmood[2,c]

[1]Physics Department, Al al-Bayt University, Mafraq 13040, Jordan
[2]Physics Department, The University of Jordan, Amman 11942, Jordan
[a]ibrahimbsoul@yahoo.com, [b]khaled_hawamda@yahoo.com,
[c]s.mahmood@ju.edu.jo

**Abstract**

$Er_3Fe_{5-x}Al_xO_{12}$ ($0.0 \leq x \leq 0.8$) garnets were prepared by ball milling and sintering at 1300°C. Rietveld refinement of the samples revealed a garnet structure with *Ia3d* symmetry. The lattice parameter, cell volume, X-ray density and magnetization of the prepared garnets decreased with the increase of Al content ($x$). The coercivity of the garnets increased with $x$, but remained generally low, being below 20 Oe. Low temperature magnetic measurements versus temperature indicated that the magnetization of $x = 0.0$ exhibited a compensation temperature at −186° C, however, $x = 0.8$ exhibited a minimum at a higher temperature of −134° C.

**Contents**

## 1. INTRODUCTION

Ferrimagnetic garnets exemplified by yttrium iron garnet (YIG) was discovery in 1956, and received considerable interest due to their low dielectric losses and remarkable performance in microwave devices and magnetic bubbles for digital memories [1, 2], in addition to their importance in the field of fundamental magnetism. The garnets have a cubic crystal structure with space group *Ia3d* and chemical formula $\{R_3^{3+}\}_c[Fe_2^{3+}]_a(Fe_3^{3+})_d\,O_{12}^{2-}$, where $R^{3+}$ is a trivalent rare-earth ion occupying dodecahedral (c) sites, and $Fe^{3+}$ ions occupy octahedral [a-sublattice] and tetrahedral (d-sublattice) sites in the garnet lattice. The magnetic properties of the garnet are determined by the strength of the superexchange interactions between magnetic ions in the various sublattices [3-6]. Since $Y^{3+}$ ion does not have magnetic moment, the magnetic properties of YIG are completely determined by a–d superexchange interactions between $Fe^{3+}$ ions at tetrahedral and octahedral sites, resulting in a net moment of 5 $\mu_B$ per molecule at 0 K. If $R^{3+}$ is a rare earth ion with net magnetic moment, however, the a–d superexchange interaction is much stronger than a–c and d–c interactions between $R^{3+}$ and $Fe^{3+}$ ions, and the rare earth sublattice couples antiferromagnetically with the net moment of the $Fe^{3+}$ sublattices. When $R^{3+}$ is a rare earth ion such as $Gd^{3+}$ through $Yb^{3+}$, the rare earth sublattice magnetization is higher than the net magnetization of the $Fe^{3+}$ sublattices, and the magnetization of the rare earth iron garnet is given by:

$$M(T) = M_c(T) - [M_d(T) - M_a(T)] \qquad (1)$$

Here the effect of $M_c$ on $M_a$ and $M_d$ through a–c and d–c interactions was neglected [1]. Since the magnetization of the rare earth sublattice decreases faster than that of the iron sublattices, the magnetization of the rare earth iron garnet as a function of temperature exhibits a compensation temperature at which the magnetization vanishes. Generally, the compensation temperature is below room temperature, and above the compensation temperature, the net magnetization of the iron sublattices *[$M_d(T) - M_a(T)$]* exceeds that of the RE sublattice, resulting in a rise of the magnetization [1].

The partial substitution of $Fe^{3+}$ ions by non-magnetic ions such as $Al^{3+}$ ions was investigated by different researchers. The substitution of $Al^{3+}$ ions in YIG was first reported to occur preferentially at tetrahedral (d) sites [7, 8], with a decreasing tendency as the concentration of $Al^{3+}$ ions increases. The results of later detailed studies indicated that the fraction of $Al^{3+}$ ions at the tetrahedral sites decreased smoothly from 1.0 at $x = 0$ to 0.6 at $x = 5$ [9, 10]. In addition, the variations of the magnetization of $Y_3Al_{1.25}Fe_{3.75}O_{12}$ with heat treatment indicated that the $Al^{3+}$ ions must have redistributed themselves among tetrahedral and octahedral sites as a result of the heat treatment [11]. More recently, room temperature (RT) Mössbauer spectra of $Y_3Fe_{5-x}Al_xO_{12}$ ($x = 0.0, 0.25, 0.5, 0.75$, and 1.0) garnets prepared by sol-gel method revealed $Fe^{3+}$ and $Al^{3+}$ cationic distribution at tetrahedral sites [12]. On the other hand, evidence of the distribution of substituted ions at octahedral (a) sites was provided based on the results of Mössbauer spectroscopy [13, 14]. On the other hand, Mössbauer spectroscopy indicated that the tetrahedral sites were the preferred sites of $Al^{3+}$ ions in substituted holmium iron garnet [15]. The reported variations of site selectivity of $Al^{3+}$ ions could be an indication of the sensitivity to the synthesis route, heat treatment, and the concentration of $Al^{3+}$ ions in the compound.

Different synthesis routes and heat treatments were adopted in the preparation of iron garnets, including sol-gel method [12, 16, 17], co-precipitation method [18], and ball milling [13, 14, 19-21]. Even though YIG garnets with different scenarios of substitution for $Fe^{3+}$, and/ or $Y^{3+}$ ions by other rare earth ions were synthesized and carefully characterized, up to our knowledge, no reports were provided on the synthesis and characterization of erbium iron garnet with partial substitution of $Fe^{3+}$ ions by $Al^{3+}$ ions. The present study is concerned with the preparation and investigation of the structural and magnetic properties of $Er_3Fe_{5-x}Al_xO_{12}$ ($0.0 \leq x \leq 0.8$) synthesized by solid state reaction. The structural refinement was used to investigate the site selectivity of $Al^{3+}$ ions, and the magnetization measurements using vibrating sample magnetometry (VSM) were used to study the effect of Al-substitution on the magnetic properties of the prepared garnets.

## 2. EXPERIMENTAL PROCEDURE

$Er_3Fe_{5-x}Al_xO_{12}$ garnets with $x = 0.0, 0.2, 0.4, 0.6$ and $0.8$ were synthesized by solid state reaction method. Stoichiometric amounts of high purity $Er_2O_3$, $Fe_2O_3$ and $Al_2O_3$ starting powders were loaded into a hardened stainless-steel cup with ball to powder ratio of 8:1, and milled at 250 rpm for 4 h. The resulting precursor was pelletized and sintered at 1300°C for 2 h. X-ray diffraction (XRD) patterns were collected using Philips PW 1720 X-ray diffractometer operating at (40 kV, 40 mA), with Cu-$K_\alpha$ radiation ($\lambda = 1.5405$ Å). The samples were scanned over the angular range $15° < 2\theta < 75°$ with $0.02°$ scanning step and speed of $1°$/min. The XRD patterns were analyzed using X'pert HighScore 2.0.1 software for phase identification, and Rietveld refinement of the crystal structure was performed using FullProf suite 2000 software. The magnetic measurements at room temperature or above were performed using Vibrating Sample Magnetometer (VSM MicroMag 3900, Princeton Measurements Cooperation), operating at a maximum applied magnetic field of ±10 kOe, whereas the magnetization measurements below room temperature were performed using a Quantum Design 9T- PPMS Ever Cool-II magnetometer.

## 3. RESULTS AND DISCUSSION

### 3.1 XRD MEASUREMENTS

The structural phases of the investigated samples were identified by X'Pert HighScore software. XRD pattern of $Er_3Fe_5O_{12}$ sample is shown in Fig. 1, and the peak positions and intensities determined by the software are shown in Fig. 2, along with the standard pattern (ICDD: 01-081-0131) for $Er_3Fe_5O_{12}$ (ErIG). It is clear from these figures that the sample with $x = 0.0$ consists of a pure EIG phase with space group Ia3d; no secondary phases were observed. Similarly, preliminary analyses of the XRD patterns of all Al-substituted samples (not shown for brevity) indicated the presence of a major garnet phase consistent with the standard pattern for ErIG.

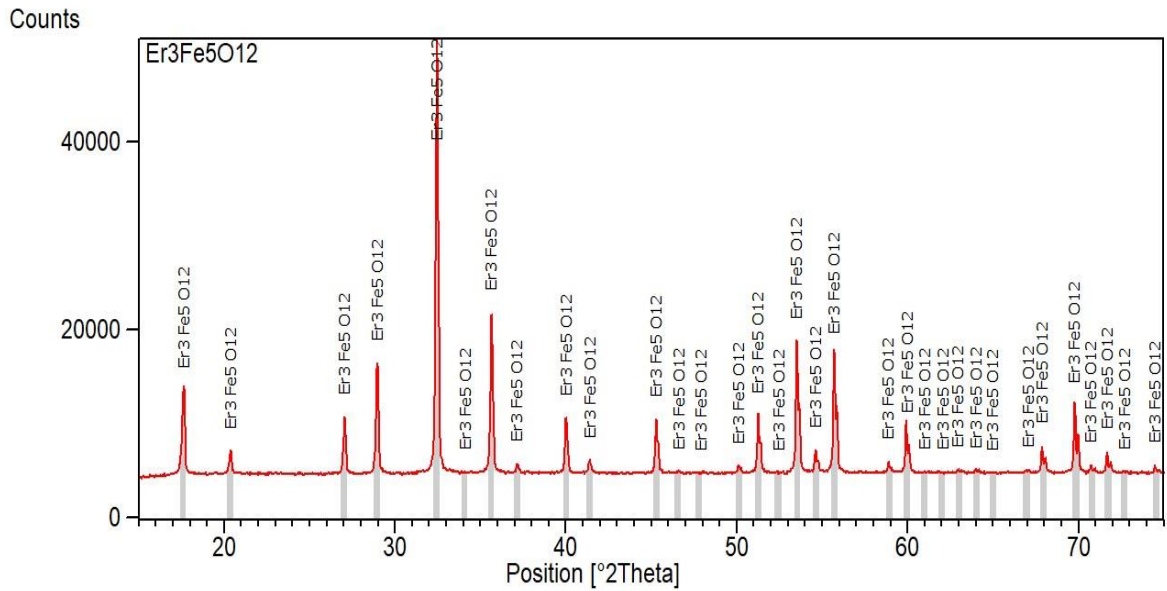

Fig.1. XRD patterns for $Er_3Fe_5O_{12}$ ($x = 0.0$) sample.

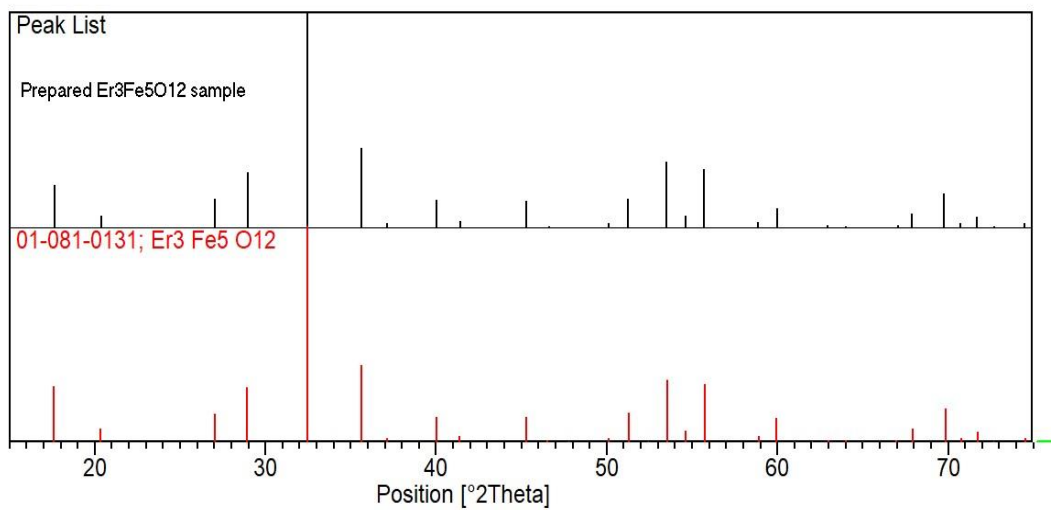

Fig. 2. XRD intensity profile of $Er_3Fe_5O_{12}$ ($x = 0.0$) sample (upper panel), and that of the standard (ICDD: 01-081-0131) pattern (lower panel).

Fig. 3 shows Rietveld refinement of the XRD patterns for all $Er_3Fe_{5-x}Al_xO_{12}$ samples, where the experimental data are represented by red circles, the theoretical pattern by black line, and the residual difference curve by blue line. The vertical (green) bars shown below the XRD patterns represent Bragg peak positions. From this figure we can see that the theoretical pattern is in good agreement with the experimental data, where the residual difference curve is a straight horizontal line with small ripples only around the main structural peak positions. The lattice constant ($a$), the unit cell volume and the density of all examined samples were obtained from the output file of the refined XRD patterns.

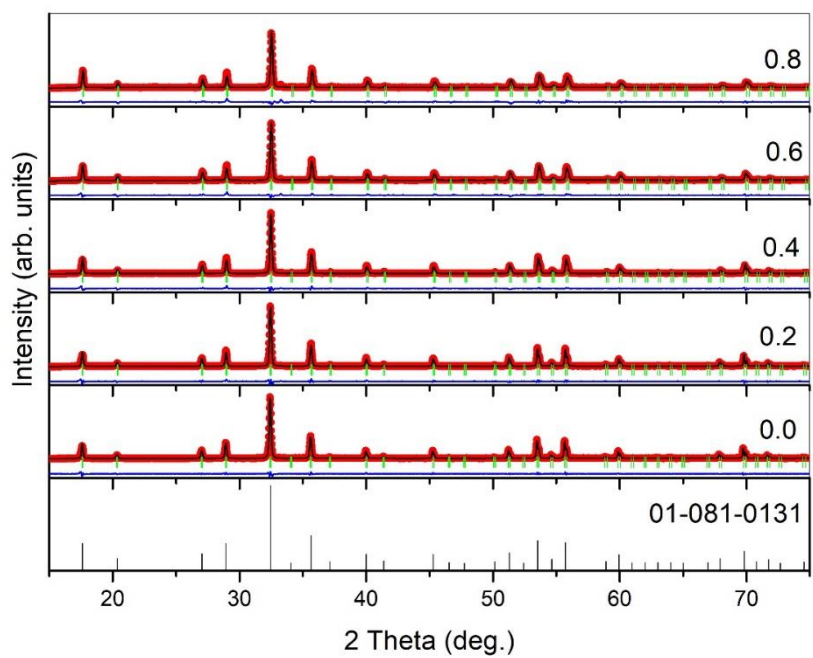

Fig.3. Rietveld refinement of the XRD patterns of all $Er_3Fe_{5-x}Al_xO_{12}$ samples.

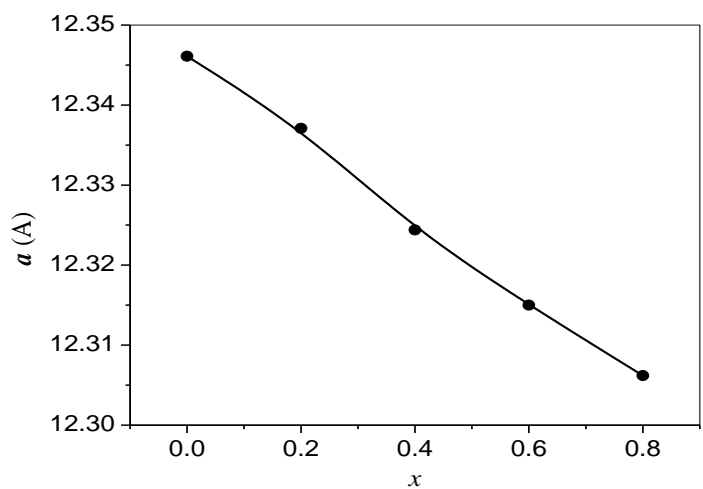

Fig.4. Lattice parameter (a) for $Er_3Fe_{5-x}Al_xO_{12}$ samples as a function of Al concentration.

Fig. 4 shows a monotonic decrease of the lattice constant ($a$) for $Er_3Fe_{5-x}Al_xO_{12}$ as $x$ increases. The lattice constant of 12.346 Å for the sample with $x = 0.0$ is in good agreement with reported value for $Er_3Fe_5O_{12}$ [1], and the observed decrease with the increase of $Al^{3+}$ content is consistent with the reported decrease of lattice constant of $Y_3Fe_{5-x}Al_xO_{12}$ with the increase of $x$ [19]. This decrease, which also results in a monotonic decrease of the unit cell volume as shown in fig. 5, is associated with the smaller ionic radius of $Al^{3+}$ ion compared to $Fe^{3+}$ ion. Also, a monotonic decrease in X-ray density was observed as $x$ increased (Fig. 6), contrary to the expected increase due to the reduction of cell volume. This is opposite to what was observed with Dy-substituted

Y in YIG [20], and can be associated with the higher rate of decrease of the molecular weight in comparison with the rate of decrease of the cell volume as $x$ increased.

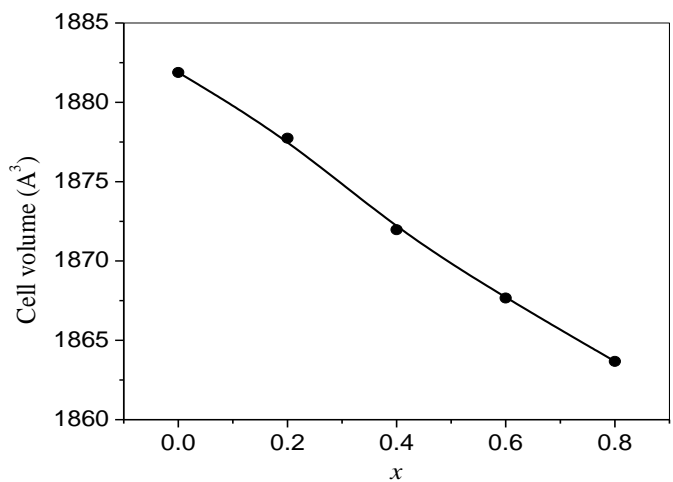

Fig. 5. Unit cell volume for $Er_3Fe_{5-x}Al_xO_{12}$ samples as a function of Al concentration.

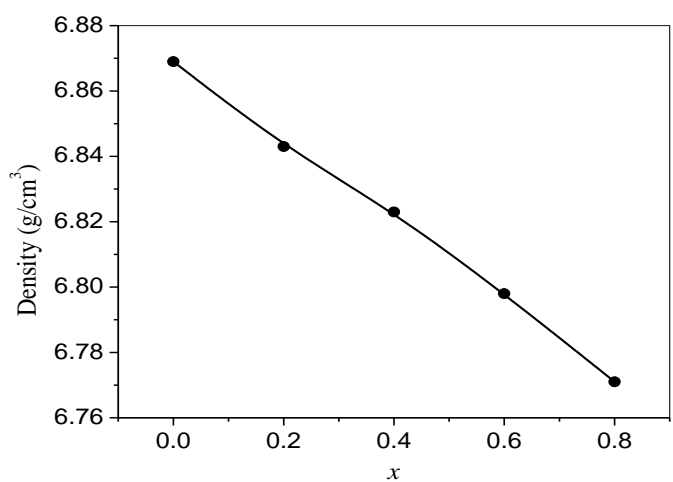

Fig. 6. Density of $Er_3Fe_{5-x}Al_xO_{12}$ samples as a function of Al concentration.

Further microstructural information was obtained from the output results (CIF file) of the Rietveld refinement of the XRD patterns. Diamond software was used to construct the crystal structure and oxygen polyhedra surrounding the metal ions, and the results for the sample with $x = 0.0$ are shown in Fig. 7. Three magnetic sublattices were observed in ErIG corresponding to magnetic ions occupying different crystallographic sites in the garnet lattice. Two of these arise from $Fe^{3+}$ ions occupying 16a (octahedral) sites and 24d (tetrahedral) sites, whereas the third corresponds to $Er^{3+}$ ions occupying 24c (dodecahedral) sites. For comparison purposes, we also constructed the unit cell (Fig. 8) for the sample $Er_3Fe_{5-x}Al_xO_{12}$ with $x = 0.8$. Similar constructions were obtained for the remaining Al-substituted samples (results are not shown to avoid redundancy). The fractions of $Fe^{3+}$ ions and $Al^{3+}$ ions at tetrahedral and octahedral sites in the investigated samples were also determined using the refinement output CIF file. The results indicated that the replacement of $Fe^{3+}$ ions by $Al^{3+}$ ions takes place only at tetrahedral

(d) sites, and that the experimental values of $Al^{3+}$ concentration in the investigated samples are in excellent agreement (within 0.04%) with the corresponding stoichiometric values in the starting powders.

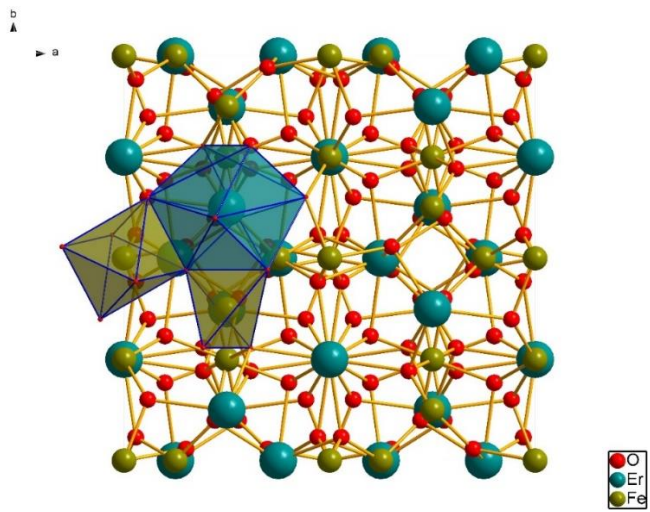

Fig.7. Crystal structure of $Er_3Fe_5O_{12}$ sample showing the polyhedra corresponding to the three magnetic sublattices in the garnet structure.

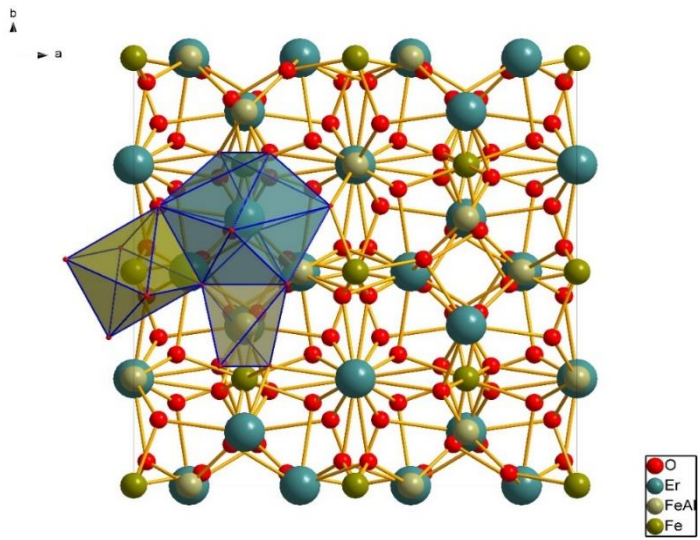

Fig.8. Crystal structure of $Er_3Fe_{4.2}Al_{0.8}O_{12}$ sample showing the polyhedra corresponding to the three magnetic sublattices in the garnet structure.

The substitution of $Fe^{3+}$ ions by $Al^{3+}$ ions resulted in small variations of the bond lengths and bond angles in the garnet structure. The trends of such variations became clear by comparing the parameters of the two extreme samples (with $x = 0.0$ and $0.8$) as shown in Table 1. The observed structural parameters of the unsubstituted ErIG sample are in good agreement with the parameters of YIG [1], even though the ionic radius of $Er^{3+}$ at dodecahedral sites (1.004 Å) is somewhat smaller than that of $Y^{3+}$ (1.019 Å) [22]. An obvious reduction of the $Fe^{3+}(d)$–$O^{2-}$ bond length from 1.851 Å for the sample with $x = 0.0$, to 1.809 Å for the sample with $x = 0.8$ was observed, whereas the $Fe^{3+}(a)$–$O^{2-}$ bond length demonstrated smaller, unsystematic variations with Al substitution. The $Fe^{3+}(a)$–$O^{2-}$–$Fe^{3+}(d)$ bond angle of 125.88° for the sample with $x = 0.0$ is in very good agreement with that reported for YIG [1, 23], and this angle increased only slightly with Al substitution. Consequently, the reduction of the tetrahedral bond length is most probably responsible for the observed monotonic decrease of the lattice constant with increasing Al concentration.

Table 1. Refined bond lengths and bond angles $Er_3Fe_5O_{12}$ and $Er_3Fe_{4.2}Al_{0.8}O_{12}$.

|  | $Er_3Fe_5O_{12}$ | $Er_3Fe_{4.2}Al_{0.8}O_{12}$ |
| --- | --- | --- |
| $Fe^{3+}(a)–O^{2-}$ (Å) | 2.023 | 2.045 |
| $Fe^{3+}(d)–O^{2-}$ (Å) | 1.851 | 1.809 |
| $Er^{3+}(c)–O^{2-}$ (Å) | 2.422 | 2.438 |
| $Fe^{3+}(a)–O^{2-}–Fe^{3+}(d)$ | 125.88° | 126.25° |
| $Er^{3+}(c)–O^{2-}–Fe^{3+}(a)$ | 101.49° | 99.85° |
| $Er^{3+}(c)–O^{2-}–Fe^{3+}(d)$ | 123.89° | 124.38° |

The crystallite size was determined from the reflection at $2\theta = 55.8°$ using Scherrer's relation. No systematic variations were observed with the increase of Al concentration, indicating that the $Al^{3+}$ substitution for $Fe^{3+}$ may not have an appreciable effect on the crystallinity of $Er_3Fe_{5-x}Al_xO_{12}$ garnets. The crystallite size was found to be $70 \pm 8$ nm for all samples, and the fluctuations around an average value, with no systematic variation, may be an indication that the microstructural characteristics did not change appreciably with Al concentration. This is consistent with the observed small, insignificant changes in bond lengths and bond angles with Al substitution.

### 3.2. MAGNETIC MEASUREMENTS

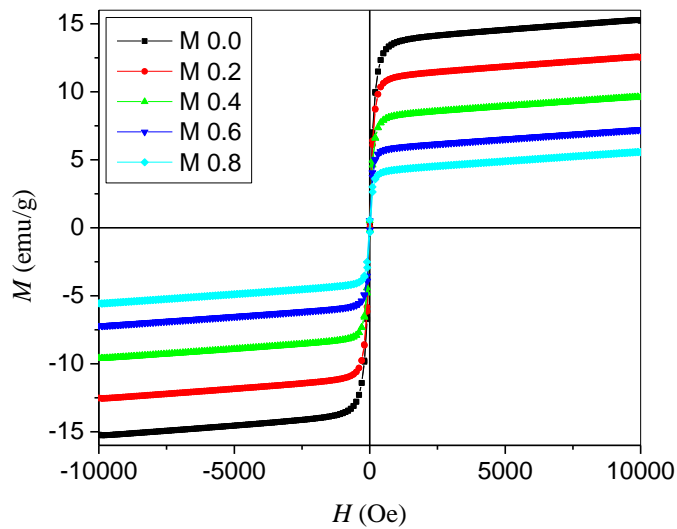

Fig.9. Hysteresis loops of $Er_3Fe_{5-x}Al_xO_{12}$ ($x = 0.0, 0.2, 0.4, 0.6, 0.8$) samples.

Fig. 9 shows the hysteresis loops for $Er_3Fe_{5-x}Al_xO_{12}$ samples measured with a 100 Oe field step. The curves revealed a soft magnetic nature of erbium iron garnets with saturation magnetization of the unsubstituted sample of 15.2 emu/g, which is in good agreement with that of 14.2 emu/g reported for ErIG [24]. The saturation magnetization decreased monotonically with the increase of Al concentration as revealed by Fig. 10, reaching a value of 5.6 emu/g at $x = 0.8$.

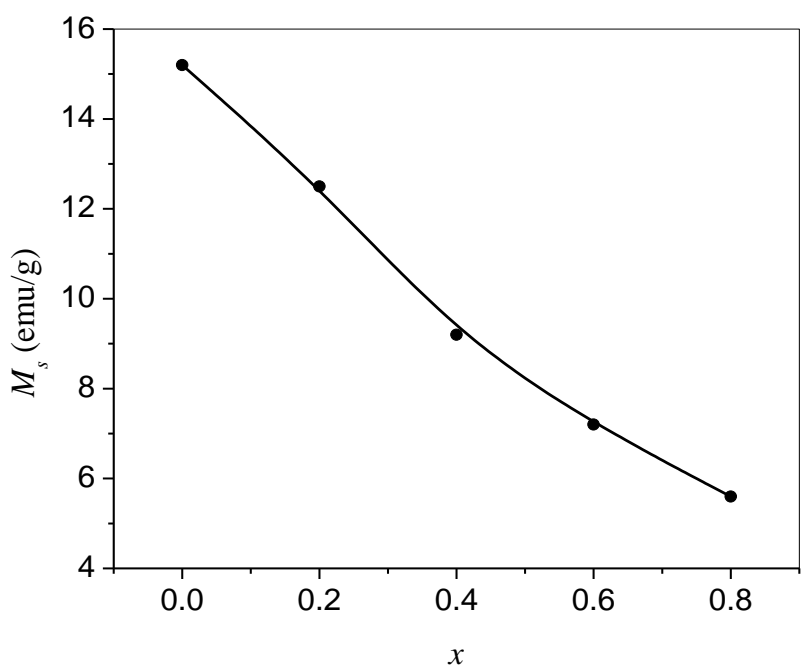

Fig.10. Saturation magnetization of $Er_3Fe_{5-x}Al_xO_{12}$ as a function of Al concentration.

The decrease of magnetization with the increase of Al content is understood if one recalls that the room temperature magnetization of ErIG is dominated by the net magnetization of the iron sublattices as mentioned in the introduction section. Accordingly, the substitution of Al at tetrahedral sites resulted in a decrease in saturation magnetization as a consequence of the decrease of the magnetization of the tetrahedral sublattice. Thus, the magnetic data supports the structural results concerning the site selectivity of $Al^{3+}$ ions.

Since the coercivity ($H_c$) of all samples was rather low, hysteresis loops in a maximum applied field of 1 kOe were measured with a 2 Oe field step, and the results are shown in Fig. 11, along with an expanded view to investigate the effect of Al substitution on the width of the loop. The coercivity of all samples was obtained directly from the hysteresis loops, and the results in Fig. 12 indicated a low coercivity of 5.8 Oe for the unsubstituted sample, and monotonic increase with the increase of Al content. However, the coercivity remained below 20 Oe for all samples.

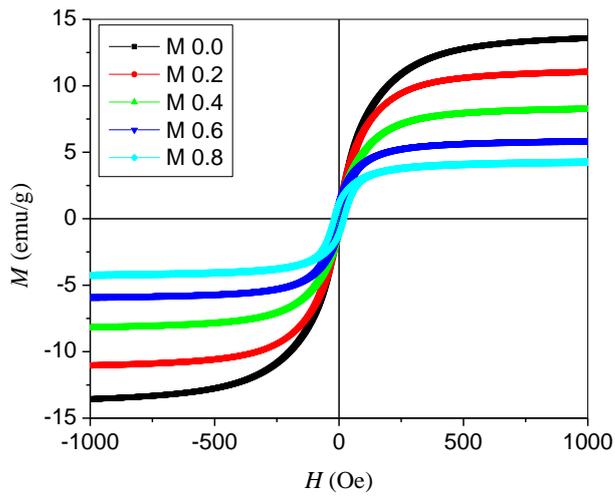

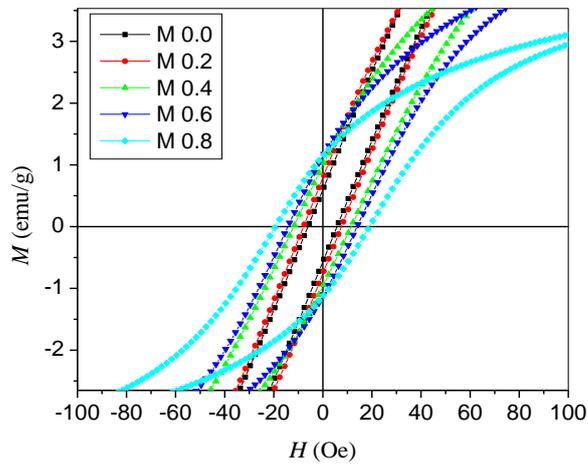

Fig.11. Hysteresis loops for $Er_3Fe_{5-x}Al_xO_{12}$ samples measured with a field step of 2 Oe.

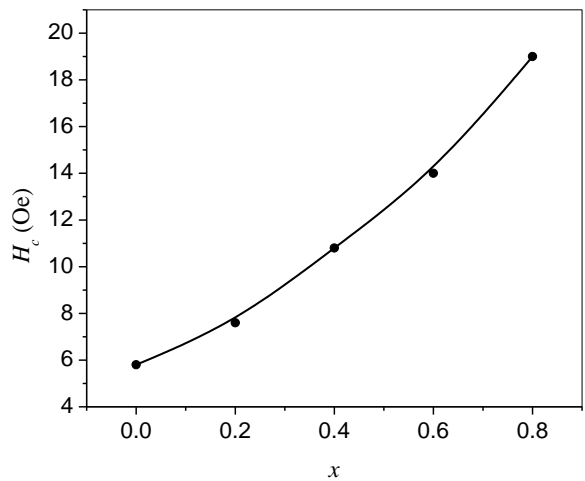

Fig.12. Coercivity as a function of Al concentration for $Er_3Fe_{5-x}Al_xO_{12}$ samples.

## 3.3. TEMPERATURE DEPENDENCE OF THE MAGNETIC PROPERTIES

The temperature dependence of the magnetization $M(T)$ was investigated by carrying out magnetization measurements as a function of temperature in a constant applied field of 8000 Oe. Fig. 13 shows magnetization curves as a function of temperature for $Er_3Fe_{5-x}Al_xO_{12}$ samples. The thermomagnetic curves demonstrated a non-linear decrease of magnetization with increasing temperature, and a sharper drop in magnetization was observed as the ferrimagnetic-paramagnetic critical transition temperature (Curie temperature; $T_c$) was approached. The critical transition temperature was obtained from the inflection point of the thermomagnetic curve revealed by the peak position in the derivative curve ($dM/dT$) as shown in Fig. 14.

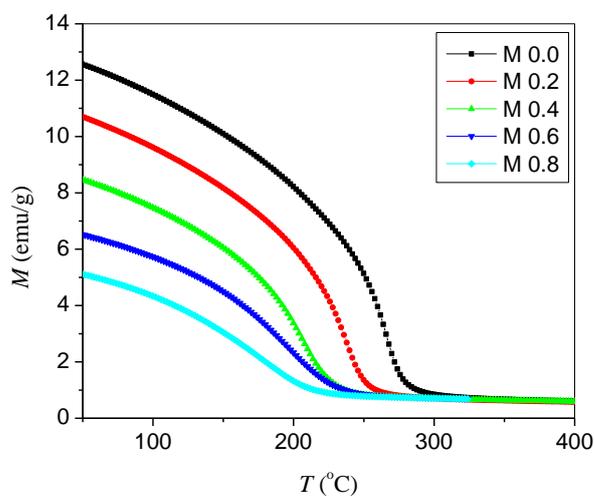

Fig.13. Magnetization as a function of temperature for $Er_3Fe_{5-x}Al_xO_{12}$ samples at a constant applied field of 8 kOe.

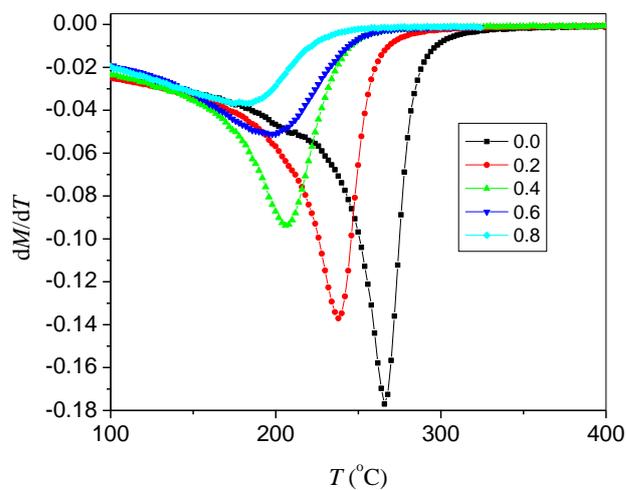

Fig.14. Derivative of the magnetization (dM/dT) of $Er_3Fe_{5-x}Al_xO_{12}$ samples as a function of temperature.

Curie temperature as a function of $Al^{3+}$ concentration in $Er_3Fe_{5-x}Al_xO_{12}$ is shown in Fig. 15, revealing a sharp decrease in $T_c$ as result of $Al^{3+}$ substitution. The structural analysis revealed a decrease of the $Fe^{3+}(d)$–$O^{2-}$ bond length, and a small increase of the $Fe^{3+}(a)$–$O^{2-}$–$Fe^{3+}(d)$ bond angle as a result of $Al^{3+}$ substitution for $Fe^{3+}$, which, contrary to observed results, should lead to an increase in the strength of the a–d superexchange interaction, and a consequent increase in $T_c$. The observed decrease of $T_c$ cannot therefore be attributed to variations of the interionic distances and bond angles, and can be associated with the weakening of the a–d superexchange interactions as a result of the decrease of the magnetization of the d-sublattice due to the replacement of magnetic $Fe^{3+}$ ions by non-magnetic $Al^{3+}$ ions in this sublattice.

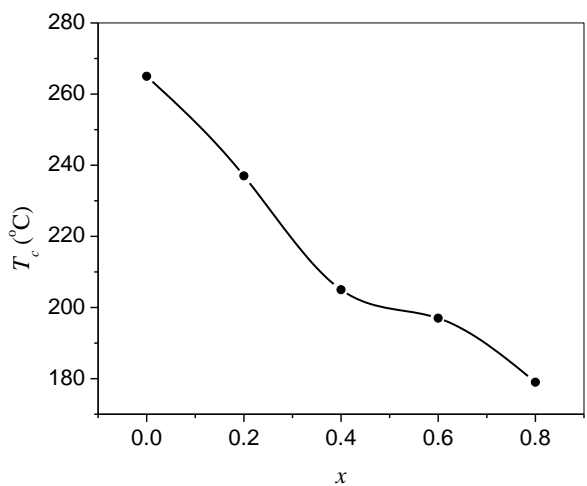

Fig.15. Curie temperature as a function of Al concentration in $Er_3Fe_{5-x}Al_xO_{12}$ samples.

### 3.4. LOW TEMPERATURE MEASUREMENTS

The temperature dependence of the magnetization $M(T)$ below room temperature was investigated by measuring the magnetization of the samples as a function of temperature in a constant applied field of 100 Oe. Fig. 16 shows the magnetization curve as a function of temperature for $Er_3Fe_5O_{12}$ sample. The magnetization decreased sharply to zero with the increase of temperature, and then increased, exhibiting a compensation temperature $T_{comp} = -186\,°C$, which in good agreement with previously reported results [25]. At temperatures below $T_{comp}$, the magnetization of the $Er^{3+}$ sublattice is larger than the net magnetization of the $Fe^{3+}$ sublattices, resulting in a net magnetization of $Er_3Fe_5O_{12}$ parallel to the magnetization of the $Er^{3+}$ sublattice. As the temperature increases, the magnetization of the $Er^{3+}$ sublattice decreases faster than that of $Fe^{3+}$ sublattices, so that the magnetization of the c-sublattice becomes equal (and antiparallel) to the net magnetization of the $Fe^{3+}$ sublattices, and the magnetization of the sample vanishes at the compensation temperature. Above $T_{comp}$, the net magnetization of the $Fe^{3+}$ sublattices continues to decrease at a slower rate compared to the magnetization of the $Er^{3+}$ sublattice, resulting in a rise in net magnetization of the sample, which becomes parallel to the net magnetization of the $Fe^{3+}$ sublattices in this temperature regime. The increase in magnetization continues with the increase of temperature, reaching a maximum value due to competition with the thermal (decreasing) effects, and then starts decreasing to zero at the Curie temperature [25].

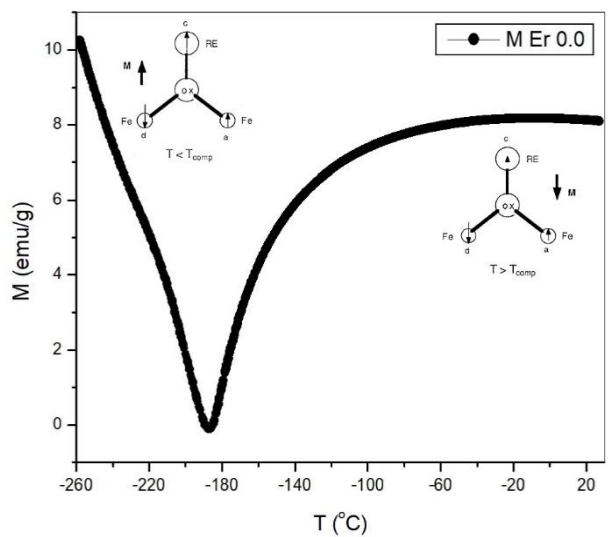

fig.16. Magnetization as a function of temperature for $Er_3Fe_5O_{12}$ sample.

The magnetization as a function of temperature for Al-substituted ErIG was also investigated in the temperature range below room temperature. In order to investigate the maximum effect of Al-substitution on the temperature dependence of the magnetic behavior of EIG, the thermomagnetic curves were measured for the two end samples with $x = 0.0$ and $0.8$, and the results are shown in Fig. 17. The magnetization curve of the sample with $x = 0.8$ exhibited a minimum at a higher temperature of $-134°$ C compared to the unsubstituted sample, with obvious flattening and broadening of the minimum peak structure. This behavior may be explained in terms of significant changes in the magnetic structure and the temperature dependence of the magnetization of the magnetic sublattices as a result of the relatively high Al substitution in the tetrahedral sublattice. At this substitution level, the net magnetization of $Fe^{3+}$ is low, which leads to the following effects: (1) The reduction of the tetrahedral sublattice magnetization reduces the strength of the a–d superexchange interaction, which may result in spin canting and a faster drop of the magnetization of the $Fe^{3+}$ sublattices with the increase in temperature. (2) Since the c-sublattice is magnetized by coupling of the rare earth magnetization with the net magnetization of $Fe^{3+}$ sublattices, the net magnetization of the c-sublattice is also expected to be lowered relative to the unsubstituted sample, and exhibit a higher degree of spin canting of the rare earth sublattice [26]. (3) This significant change in magnetic structure should, in principle, influence the temperature dependence of the magnetization of the magnetic sublattices. The higher magnetization of the Al-substituted sample relative to the unsubstituted sample, even up to temperatures exceeding the compensation temperature of the unsubstituted sample, is an indication of the dominance of the rare earth magnetic sublattice over a wider temperature range in the Al-substituted sample. This increase in magnetization in the low temperature range is a result of the significant reduction of the $Fe^{3+}$ sublattice magnetization due to the depletion of $Fe^{3+}$ ions at the tetrahedral site. Also, it was observed that the net magnetization of the Al-substituted sample did not vanish completely at the point of minimum magnetization, which requires further investigation.

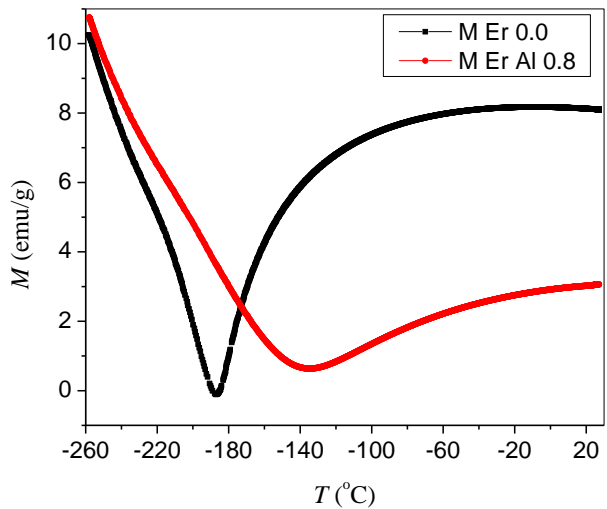

Fig.17. Magnetization as a function of temperature for $Er_3Fe_{5-x}Al_xO_{12}$ ($x = 0.0$ and $0.8$) samples.

## 4. CONCLUSIONS

X-ray diffraction of the Al-substituted $Er_3Fe_{5-x}Al_xO_{12}$ ($x = 0.0$ to $0.8$) samples prepared by ball milling and sintering at 1300º C indicated crystallization of the garnet phase with decreasing lattice parameter, cell volume, and X-ray density as a result of the increase of Al-substitution. The magnetization decreased significantly with the increase of Al content, and the coercivity increased, while remaining below 20 Oe for all samples. The Curie temperature also decreased appreciably with the increase of Al content, which is an indication of the reduction of the strength of superexchange interaction resulting from the reduction of the tetrahedral moment as a consequence of Al substitution. The unsubstituted sample exhibited the usual compensation point of erbium iron garnet, while the thermomagnetic curve of the sample with $x = 0.8$ exhibited a minimum at a higher temperature, with significant broadening. This could be an indication that Al substitution results in a significant change of the temperature dependence of the magnetic sublattices, in addition to the reduction of the overall magnetization above the compensation point of EIG.